\input{aipcheck}

\documentclass[
    ,final            
  ]
  {aipproc}

\layoutstyle{6x9}

\begin{document}

\title{Characterizing the magnetic fields of the first $\tau$\,Sco analogues}


\author{V. Petit}{
  address={West Chester University, USA}
}

\author{O. Kochukhov}{
  address={Uppsala University, Sweden}
}

\author{D.~L. Massa}{
  address={Space Telescope Science Institute, USA}
}

\author{W.~L.~F. Marcolino}{
  address={Universidade Federal do Rio de Janeiro, Brazil}
}

\author{G.~A. Wade}{
  address={Royal Military College of Canada, Canada}
}

\author{R. Ignace}{
  address={East Tennessee State University, USA}
}

\begin{abstract}
 The B0.2 V magnetic star $\tau$\,Sco stands out from the larger population of massive OB stars due to its high X-ray activity, peculiar wind diagnostics and complex magnetic field. Recently, Petit et al. \cite{Petit2011} presented the discovery of the first two $\tau$\,Sco analogues -- HD\,66665 and HD\,63425, identified by the striking similarity of their UV spectra to that of $\tau$\,Sco. ESPaDOnS and Narval spectropolarimetric observations were obtained by the Magnetism in Massive Stars CFHT and TBL Large Programs, in order to characterize the stellar and magnetic properties of these stars. A magnetic field of similar surface strength was found on both stars, reinforcing the connection between the presence of a magnetic field and wind peculiarities. We present additional phase-resolved observations secured by the MiMeS collaboration for HD\,66665 in order to measure its magnetic geometry, and correlate that geometry with diagnostics of mass-loss.\end{abstract}

\maketitle


\section{The Young Magnetic B-type star $\tau$\,Sco}

The B0.2 V star $\tau$\,Sco is recognised to be a peculiar and outstanding object:

1. The magnetic field of $\tau$\,Sco is unique because it is structurally more complex than the mostly-dipolar fields (l=1) usually observed in hot magnetic OB stars, with significant power in spherical-harmonic modes up to l=5 with a mean surface field strength of $\sim$300 G (Donati et al. \cite{Donati2006}).

2. $\tau$\,Sco also stands out from the crowd of early-B stars because of its stellar wind anomalies, as diagnosed through its odd UV spectrum that lies outside the normal temperature-luminosity classification grid (Walborn, Parker \& Nichols \cite{Walborn1995}). The hard X-ray emission of $\tau$\,Sco also suggests hot plasma (Cohen et al. \cite{Cohen2003}).

Interestingly, the wind lines of $\tau$\,Sco vary periodically with the star's 41\,d rotation period (Donati et al. \cite{Donati2006}). Clearly the magnetic field exerts an important influence on the wind structure. What is not clear is whether the wind-line anomalies described above are a consequence of the unusual complexity of $\tau$\,Sco's magnetic field, a general consequence of wind confinement in this class of star, or perhaps even unrelated to the presence of a magnetic field.
Because such wind anomalies have never been observed in any other star, magnetic or not, this issue has remained unresolved.

\section{The $\tau$\,Sco analogues}

\begin{figure}
  \includegraphics[height=0.5\textheight]{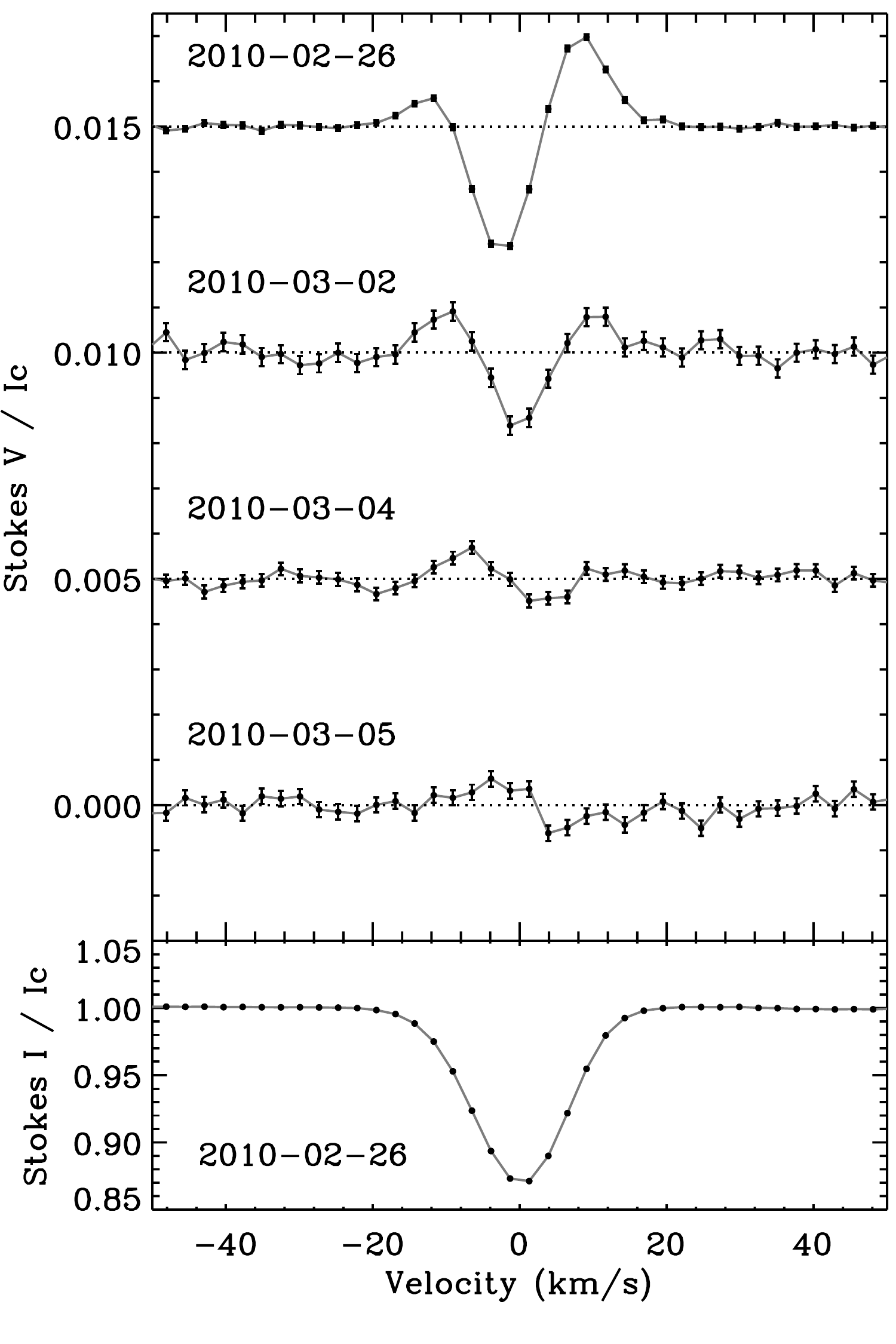} \includegraphics[height=0.5\textheight]{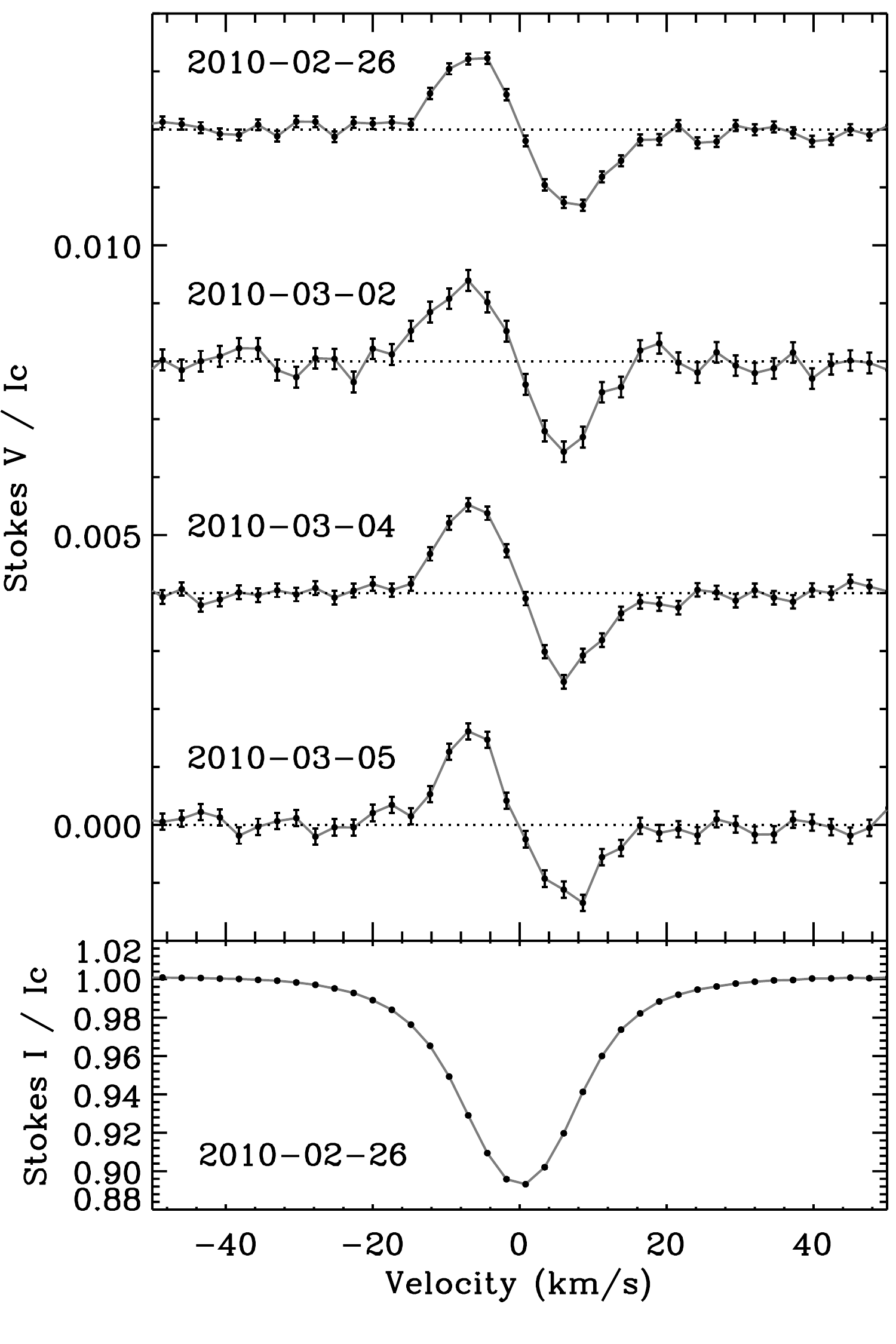}
  \caption{\label{fig_lsd}Mean Stokes I absorption line profiles (bottom) and circular polarisation Stokes V profiles (top) of HD\,66665 (left) and HD\,63425 (right) obtained with ESPaDOnS. All LSD profiles have been scaled to correspond to a spectral line with weight corresponding to $d=0.2\,I_c$, $g=1.2$ and $\lambda_0=5000$\,\AA.}
\end{figure}

In this poster we presented two early B-type stars -- HD\,66665 and HD\,63425 -- that we identified to be the first $\tau$\,Sco analogues. These stars were first discovered by their UV spectra, which are strikingly similar to the UV spectrum of $\tau$\,Sco. The discovery of wind anomalies naturally led to an investigation of the magnetic properties of these stars (Petit et al. \cite{Petit2011}) by the Magnetism in Massive Stars collaboration (Wade et al. \cite{Wade2010}).
Spectropolarimetric observations of HD\,66665 and HD\,63425 were taken with ESPaSOnS at the Canada-France-Hawaii Telescope. We first acquired 4 high-resolution, broad-band, intensity (Stokes I) and circular polarisation (Stokes V) spectra for each star.

In order to determine the stellar and wind parameters of HD\,66665 and HD\,63425 we used non-LTE model atmospheres from the \textsc{cmfgen} code (Hillier \& Miller \cite{Hillier1998}). The apparent rotational velocity of these two stars is clearly low, but an accurate value is difficult to infer from the observed spectrum, due to turbulent broadening of comparable or larger magnitude.

UV spectra from IUE were used to infer stellar wind properties. The absorption shown by C\textsc{iv} cannot be reproduced in detail by our models. On the other hand, we could achieve an acceptable fit for the S\textsc{iv} and N\textsc{v} lines. The inclusion of X-rays in our \textsc{cmfgen} models was essential in reproducing the N\textsc{v} feature.

We have found that $\tau$\,Sco and the two new analogues have similar fundamental properties. However, the mass-loss rates we estimate for HD\,66665 and HD\,63425 are lower than the value adopted for $\tau$\,Sco by Donati et al. \cite{Donati2006} in their analysis of that star. However, that mass-loss rate (Mokiem et al. \cite{Mokiem2005}) was determined from the optical spectrum of $\tau$\,Sco and could differ systematically from that determined from UV line profiles (e.g. Oskinova et al. \cite{Oskinova2011}) -- in particular in the presence of a magnetic field. 

\begin{figure}
  \includegraphics[height=0.5\textheight]{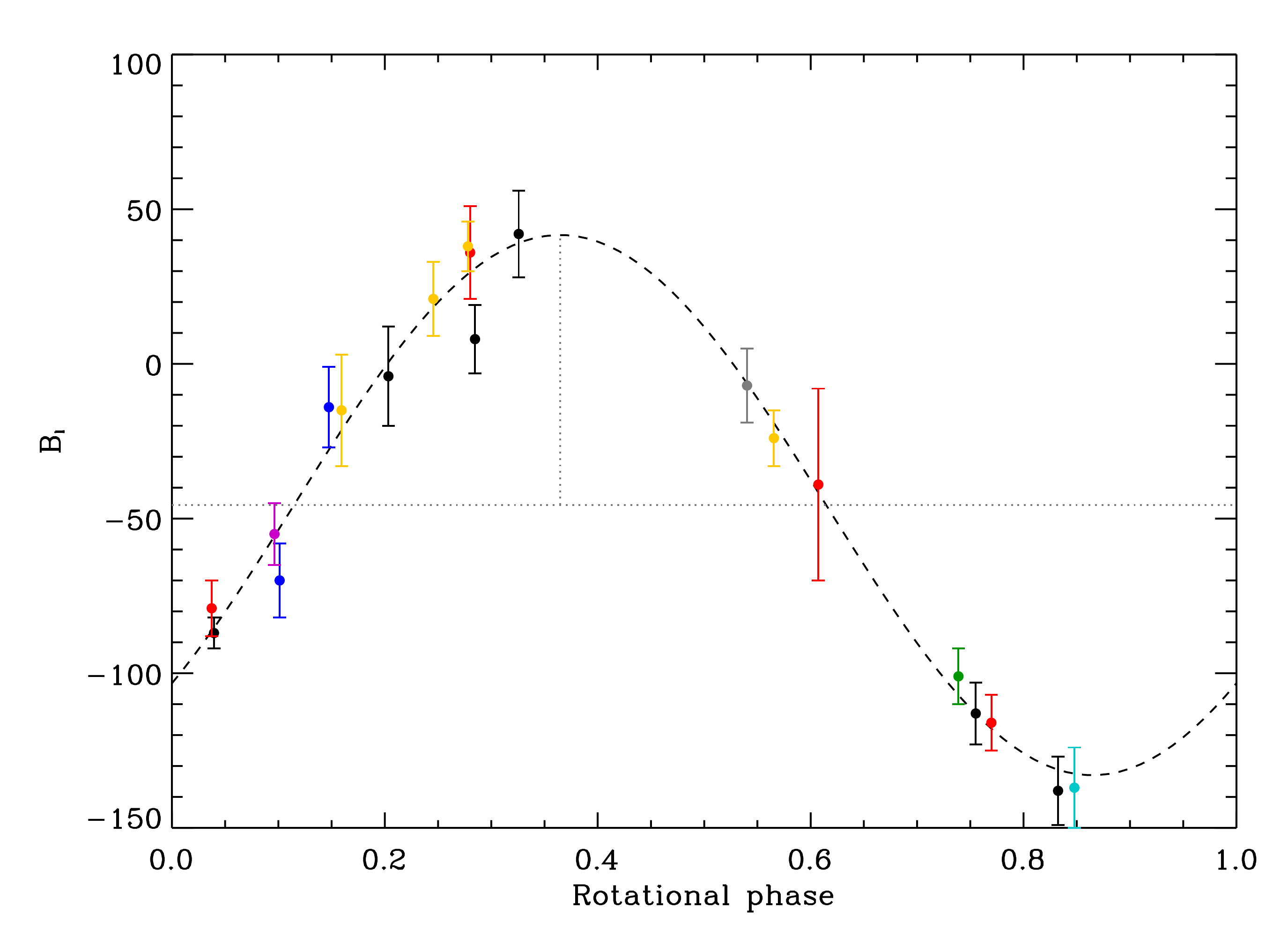}
  \caption{\label{bl}Disk-integrated longitudinal field curve for HD\,66665, phased with a period around 20 days. The dash curve shows a sinusoidal fit. }
\end{figure}

In order to increase the magnetic sensitivity of our data, we applied the LSD procedure, which enables the simultaneous use of many spectral lines to detect a magnetic field Stokes V signature (Donati et al. \cite{Donati1997}; Kochukhov et al. \cite{Kochukhov2010}). All observations led to the detection of a magnetic signal (see Figure \ref{fig_lsd}). The same analysis was performed on the diagnostic null profiles, and no signal was detected.
As the exact rotation phases of our observations were not known, we used the method described by Petit et al. \cite{Petit2008}, which compares the observed Stokes V profiles to a rotation independent, dipolar oblique rotator model, in a Bayesian statistic framework. From this we obtained a conservative estimate of the dipolar surface field strength.

The average surface field strengths are of the same scale as the surface field of $\tau$\,Sco. However, more phase-resolved observations were required in order to assess the potential complexity of their magnetic field, and verify if the wind anomalies are linked to the field complexity.

\section{New observations of HD\,66665}

We monitored HD\,66665 with the spectropolarimeters ESPaDOnS and Narval. A total of 20 observations were acquired between March 2010 and April 2011. The magnetic measurements can be phased with a period around 20 days. Figure \ref{bl} shows the folded longitudinal field curve. Although the phase coverage is not complete enough to attempt magnetic Doppler imaging of the surface field (Piskunov \& Kochukhov \cite{Piskunov2002}), the global longitudinal field variation is much smoother than $\tau$\,Sco's and can be fit by a sinusoidal curve (shown by the dashed curve). The magnetic field of HD\,66665 probably does not differ from a dipole configuration as much as $\tau$\,Sco's does.

\begin{theacknowledgments}
Based on observations obtained at the Canada-France-Hawaii Telescope (CFHT) and at the T\'elescope Bernard Lyot (TBL). CFHT is operated by the National Research Concil of Canada, the Institut National des Sciences de l'Univers of the Centre National de la Recherche Scientifique of France, and the University of Hawaii. TBL is operated by CNRS/INSU.
\end{theacknowledgments}

\bibliographystyle{aipproc}

\end{document}